\documentclass[amsmath, amssymb,10pt, aps, prb, twocolumn, notitlepage, longbibliography, superscriptaddress]{revtex4-1}
\usepackage{graphicx}
\usepackage{calc}
\usepackage{braket}
\usepackage{ulem}   % to strike things out
\usepackage{slashed}
\usepackage{wasysym}
\usepackage{dsfont}
\usepackage{amsthm,amsmath,amsfonts,amssymb,verbatim,color}
\usepackage{lipsum}
\usepackage{subfigure}
\usepackage{bm}
\usepackage{epsfig,slashed}
\usepackage[T1]{fontenc}
\usepackage{hyperref}
\usepackage{flafter}

\newcommand{\bj}{{\bf j}}
\newcommand{\bk}{{\bf k}}
\newcommand{\bP}{{\bf P}}
\newcommand{\bB}{{\bf B}}
\newcommand{\bE}{{\bf E}}

\def\bnabla{{\bm \nabla}}
\def\hbsigma{\hat{\boldsymbol \sigma}}

\newcommand{\bp}{\mathbf{p}}
\newcommand{\nim}{n_{\rm imp}}
\newcommand{\bpp}{\mathbf{p}^\prime}

\newcommand{\ve}{\varepsilon}

\newcommand{\be}{\begin{equation}}
\newcommand{\ee}{\end{equation}}

\begin{document}
\title{Adiabatic dechiralisation and thermodynamics of Weyl semimetals}

\author{S.V.~Syzranov}
\affiliation{Physics Department, University of California, Santa Cruz, CA 95064, USA}

\author{Ya.I.~Rodionov}
\affiliation{Institute for Theoretical and Applied Electrodynamics,Russian Academy of Sciences, Izhorskaya Str. 13, Moscow, 125412 Russia}
\affiliation{National University of Science and Technology MISIS, Moscow, 119049 Russia}

\author{B.~Skinner}
\affiliation{Department of Physics, Massachussetts Institute of Technology, Cambridge, MA 02139, USA}
\affiliation{Physics Department, University of California, Santa Cruz, CA 95064, USA}
\date{\today}

%%%%%%%%%%%%%%%%% END OF PREAMBLE %%%%%%%%%%%%%%%%

\begin{abstract}

We study thermodynamic manifestations of the chiral anomaly in disordered Weyl semimetals.
We focus, in particular, on the effect which we call "adiabatic dechiralisation", the phenomenon in which a change in temperature and/or an absorption or release of heat results from applying parallel electric and magnetic fields that change the imbalance of quasiparticles with different chiralities (at different Weyl nodes).
This effect is {similar} to that of adiabatic demagnetisation, which is commonly used as a method of low-temperature refrigeration.
 We describe this phenomenon quantitatively
and discuss experimental conditions favourable for its observation.
A related phenomenon, which we analyse and which is readily observable in experiment, is the
dependency of the heat capacity
of a Weyl
semimetal on parallel electric and magnetic fields.
\end{abstract}

\maketitle

Weyl semimetals are solid-state systems with Weyl quasiparticle dispersion near certain points (the Weyl nodes)
in (quasi)momentum space~\cite{ArmitageAshvin:review}.
One of the most fascinating features of Weyl semimetals is the chiral anomaly\cite{Burkov:review,SonSpivak:anomaly,Parameswaran:probingAnomaly,Burkov:AnomalyDiffusive},
the transfer of quasiparticles from one Weyl node to another when external electric and magnetic fields
are applied simultaneously. This phenomenon is the solid-state equivalent of the anomalous non-conservation
of the chiral current\cite{Adler:anomaly,BelJackiw:anomaly} which has been predicted for elementary particles with Weyl dispersion.

The chiral anomaly has been predicted~\cite{SonSpivak:anomaly} to lead to a negative longitudinal magnetoresistance in Weyl semimetals,
with the corresponding correction $\Delta \sigma_\parallel$ to the conductivity growing as $\propto B^2$ in the limit of small magnetic fields $B$,
which results from the transfer of quasiparticles from one node to another.
Observing this negative longitudinal magnetoresistance with a strong directional dependence
has so far been the main focus of experimental studies
of the chiral anomaly in Weyl semimetals (see, e.g., Refs.~\onlinecite{Huang:negativeMR,ZHasan:AnomalySignatures,Yang:negativeMR, JianHua:negativeMR,Wang:negativeMR,Jia:ReviewMR, Xiong:evidence, Hirschberger:chiralanomaly,Ong:CurrentJettingClaim}).
Unfortunately, such transport studies are often hindered by a number of magnetoresistance phenomena that are unrelated to the chiral anomaly. For example, a well-known experimental issue is ``current jetting'', in which a strong magnetic field causes the current to focus tightly along the field direction
\cite{dosReis:chiralanomaly, Arnold:negativeMR, Hirschberger:chiralanomaly}. This effect can give rise to a negative magnetoresistance in experiment that resembles the chiral anomaly but has no chiral origin. At the same time,
a number of positive magnetoresistance
effects~\cite{Abrikosov:metals,Dreizin:largeB,Song:linearMR,Murzin:MR} appear generically in disordered three-dimensional conductors and
may mask the effect of the chiral anomaly.

These difficulties motivate us to consider thermodynamic manifestations of the chiral anomaly in Weyl semimetals.
In particular, in this paper we explore the phenomenon of ``{\it adiabatic dechiralisation}'' (which we call so by analogy
with {adiabatic demagnetisation~\cite{Klerk:demagnetization}}),  in which changing the chirality imbalance of a Weyl semimetal through applied electric and magnetic fields leads to a change in temperature and/or an absorption or release of heat.
A related thermodynamic manifestation of the chiral anomaly, readily observable in experiment, is the dependence of the heat capacity
of a Weyl on parallel electric and magnetic fields. These effects may be used as thermodynamic litmus tests to identify systems
with Weyl quasiparticle dispersion.

\begin{figure}[ht!]
	%\begin{center}
		\includegraphics[width=1.0\columnwidth]{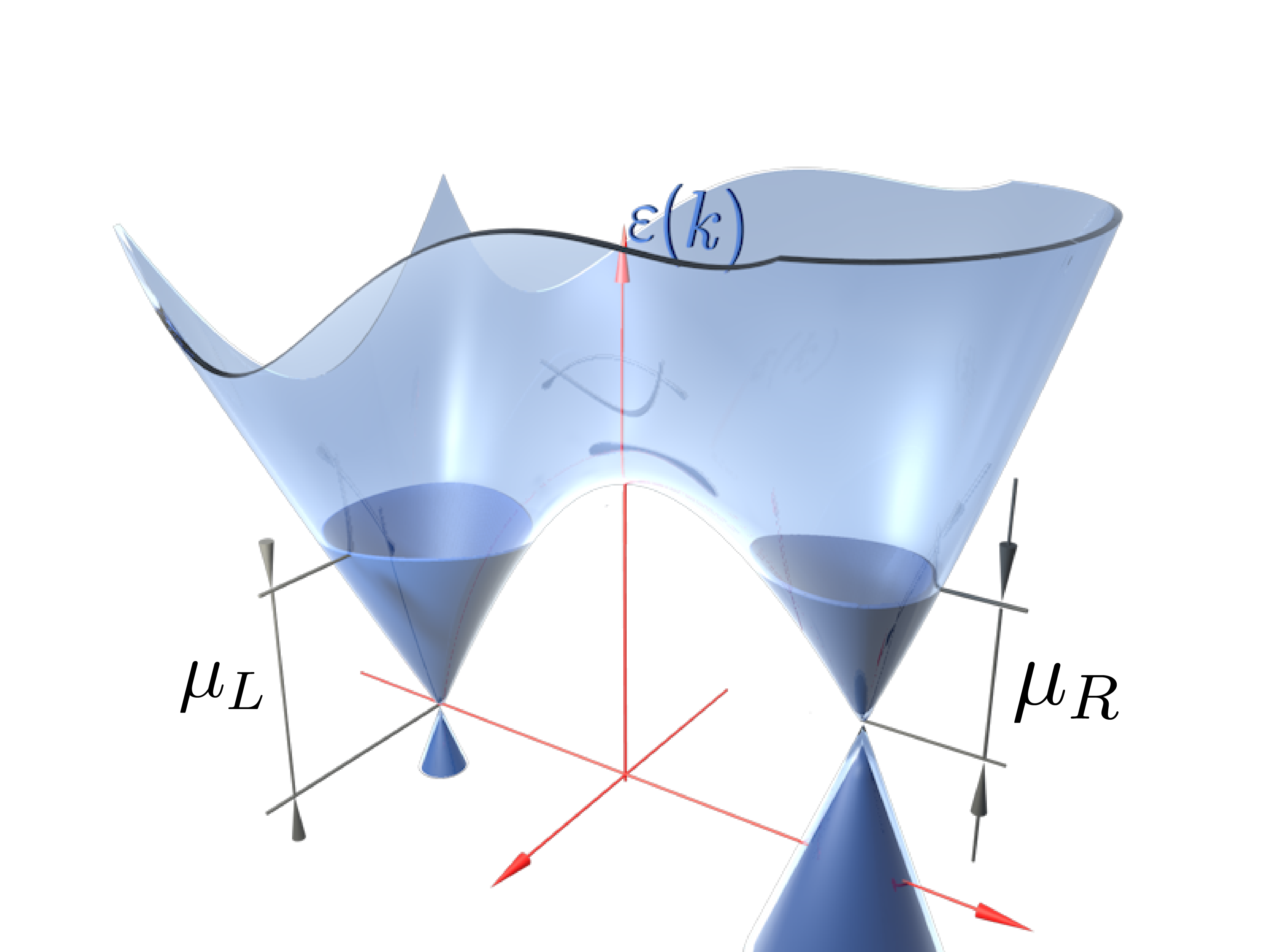}
	%\end{center}
	\caption{\label{Fig:cones}
		(Colour online)
		The band structure of a Weyl semimetal with two Weyl nodes ($L$ and $R$).
		The chemical potentials $\mu_L$ and $\mu_R$ at the nodes are measured
		from the energies of the nodes and, in general, are different.
	}
\end{figure}

{\it Model.} According to the fermion-doubling theorem~\cite{NielsenNinomiya},
a Weyl semimetal has necessarily an even number of Weyl nodes.
For simplicity, we consider a model of a Weyl semimetal with only two nodes, which we label
left ($L$) and right ($R$), with the same Fermi velocity $v$.
Although most Weyl semimetals studied experimentally have several identical Weyl cones,
Weyl systems lacking inversion symmetry are expected~\cite{Zyuzin:WSMinversionBreaking} in general
to have nodes with different energies; this is the case, for example, in SrSi$_2$~\cite{ZHasan:SrSi2}.
In this paper, we assume that the chemical potentials $\mu_L$ and $\mu_R$ of quasiparticles at the two nodes, measured from
the respective energies of the nodal points (as shown in Fig.~\ref{Fig:cones}) may have arbitrary values.
We assume also that nodes $L$ and $R$ have, respectively, negative and positive
chiralities, i.e.\ they are described by the Hamiltonians $\hat H_L = -v \hbsigma\cdot \hat\bP+\Delta_L$
and $\hat H_R = + v \hbsigma \cdot \hat\bP+\Delta_R$, where $\hbsigma$ is a pseudospin operator, $\hat\bP$ is the operator of momentum
measured from the node, and $\Delta_L$ and $\Delta_R$ are constants.

The chemical potentials in realistic Weyl semimetals are determined by doping by
donor and acceptor impurities.
When only one type of impurity (donors or acceptor) is dominant, such doping
leads generically to a homogeneous Fermi sea near each node, whose depth $\mu_i$ significantly exceeds the
characteristic fluctuations of the disorder potential arising from the randomness of impurity locations~\cite{RodionovSyzranov:impurityWeyl}.
In symmetric Weyl semimetals (with $\mu_L=\mu_R$ in equilibrium), it is possible, in principle, to bring
the chemical potential close to both nodal points, e.g.\ by balancing the concentrations
of donors and acceptors. In that case, the disorder potential fluctuations become
larger than the typical Fermi energy, leading to the formation of electron
and hole puddles\cite{Skinner:minimalCond}. While our main focus is on systems with homogenenous Fermi seas at the nodes,
our results may also be applied qualitatively to the ``puddled'' Weyl semimetals of the latter type.
In realistic systems, the chemical potentials (typically several dozen meV) are on the order of or
exceed the temperature, with the only possible exception\footnote{\label{Exception}
Recently, power-law temperature dependencies of conductivity were reported in certain iridate compounds
	in Ref.~\onlinecite{SleightRamirez:T4},
	consistent with existence of Weyl nodes with zero chemical potentials
	in the presence of charged impurities~\cite{DasSarma:T4,RodionovSyzranov:impurityWeyl}.}
 reported recently in Ref.~\onlinecite{SleightRamirez:T4}.

We assume also that the processes of quasiparticle transfer between nodes
are slow compared to the intranodal equilibration. In this limit the system may always be assumed to be in a
quasiequilibrium state, with each node $i$ having a well defined
chemical potential $\mu_i$. We emphasise that quasiparticles
at different nodes interact with each other,
and their relatively fast equilibration allows us to assume that both nodes have the same temperature $T(t)$ at each moment of time $t$.

{\it Internodal charge dynamics.} In the presence of external electric $\bE$ and magnetic $\bB$
fields, the dynamics of the concentrations $N_L$ and $N_R$ of electrons at the nodes
of a realistic Weyl semimetal are described by linear rate equations (derived microscopically
in the Supplemental Material~\cite{Supplemental}):
\begin{align}
	\frac{d N_L}{dt}=-\frac{dN_R}{dt}=\frac{g e^2}{4\pi^2 \hbar^2 c}\bE\cdot\bB
	-\frac{\delta N_L}{\tau_{L\rightarrow R}}+\frac{\delta N_R}{\tau_{R\rightarrow L}},
	\label{NDynamics1}
\end{align}
where $\delta N_i=N_i-N_i^0=\int d\varepsilon\, \nu_i(\varepsilon)\left[f(\varepsilon,\mu_i)-f^0(\varepsilon)\right]$ is the deviation of the concentration of electrons at node $i$ ($i=L,R$)
from its equilibrium value;
$f(\varepsilon,\mu_i)$ and $\nu_i(\varepsilon)$ are the distribution function
and the density of states of the quasiparticles at the $i$-th node;
$f^0(\varepsilon)$ is the equilibrium distribution function;
$g$ accounts for the
spin and possible additional valley degeneracy;
$\tau_{L\rightarrow R}^{-1}$ and $\tau_{R\rightarrow L}^{-1}$ are, respectively, the
rates of qusiparticle scattering from node $L$ to node $R$ and vice versa due to impurity scattering
(derived microscopically in the Supplemental Material~\cite{Supplemental}).
The first term on the right-hand side of Eqs.~(\ref{NDynamics1}) describes the chiral anomaly.
In Eqs.~(\ref{NDynamics1}) we used the smallness of the deviations $\delta N_i$ to linearise
the rates of change of the concentrations due to impurity collisions; in general, the rate equations are non-linear in $\delta N_i$~\cite{Supplemental}.

The concentration $N_i$ of electrons near node $i$ may in principle be defined relative to an arbitrary constant.
It is convenient to measure the concentration relative to its value when the chemical potential is at the respective nodal point ($\mu_i=0$), which gives\cite{RodionovSyzranov:impurityWeyl}
\begin{align}
	N_i=g\frac{\mu_i^3+\pi^2\mu_i T^2}{6\pi^2 v^3\hbar^3}.
	\label{Ni}
\end{align}
The conservation of the total number of electrons requires that $\delta N_L=-\delta N_R$, which is consistent with Eq.~(\ref{NDynamics1}).

Applying stationary electric and magnetic fields to the system leads, at times exceeding the internodal scattering times,
$t\gg \left(\tau_{L\rightarrow R}^{-1}+\tau_{R\rightarrow L}^{-1}\right)^{-1}$ [which in realistic systems
is on the order of dozens of picoseconds (see, e.g., Ref.~\onlinecite{ZHasan:AnomalySignatures})],
to non-equilibrium stationary concentrations of the charge carriers with
\begin{align}
	\delta N_L=-\delta N_R=\frac{ge^2}{4\pi^2\hbar^2 c}\frac{\bE\cdot\bB}{\tau_{L\rightarrow R}^{-1}+\tau_{R\rightarrow L}^{-1}}.
	\label{deltaN}
\end{align}
The modification of the stationary concentrations $N_i$ results in detectable changes of thermodynamic observables, such
as energy, entropy and heat capacity.

{\it Entropy in the stationary state.} The entropy
of a Weyl semimetal (per unit volume) in a quasiequilibrium state
%with the chemical potentials
%of the nodes $\mu_i$
is given by\cite{Supplemental}
\begin{align}
	S=\sum_{i=L,R}g\frac{7\pi^2T^3+15\mu_i^2 T}{90v^3\hbar^3}.
	\label{Sgeneric}
\end{align}
External electric and magnetic fields shift the chemical potentials $\mu_i$ from their equilibrium values,
with the deviations $\delta\mu_i$ related to the concentration changes $\delta N_i$ by
\begin{align}
	\delta \mu_i\left(\delta N_i\right)
	=
	&\frac{6\pi^2v^3\hbar^3}{g(3\mu_i^2+\pi^2T^2)}\delta N_i
	\nonumber\\
	&-\frac{216\pi^4 v^6\hbar^6\mu_i}{g^2(3\mu_i^2+\pi^2T^2)^3}\delta N_i^2+\ldots,
	\label{MuNi}
\end{align}
as follows from Eq.~(\ref{Ni}).
In terms of the concentration changes $\delta N_L=-\delta N_R$, the entropy of the system in the
experimentally important case of low temperatures $T\ll\mu_L,\mu_R$
is given by
\begin{align}
	S=S_0
	&+\frac{2\pi^2}{3} \left(\frac{T}{\mu_L}-\frac{T}{\mu_R}\right)\delta N_L
	\nonumber\\
	&-\frac{2\pi^4 v^3\hbar^3 T}{3g}\left(\frac{1}{\mu_L^4}+\frac{1}{\mu_R^4}\right)\delta N_L^2+\ldots,
	\label{SdeltaN}
\end{align}
where $S_0$ is the entropy in the absence of external fields, given by Eq.~(\ref{Sgeneric}) with the chemical potentials $\mu_i$ set to their equilibrium values.

Assuming that the deviations of the concentrations $\delta N_L=-\delta N_R$ caused by external electric and magnetic
fields are stationary, and
utilising Eqs.~(\ref{MuNi}) and (\ref{Sgeneric}), we arrive at an expression for the entropy of a Weyl semimetal
of the form
\begin{align}
S=S_0- A_1\cdot T(\bE\cdot\bB)-A_2\cdot T(\bE\cdot\bB)^2+\ldots,
\label{SEB}
\end{align}
where the coefficients $A_1$ and $A_2$ are given by
\begin{align}
A_1&=\frac{ge^2}{6\hbar^2 c\left(\tau_{L\rightarrow R}^{-1}+\tau_{R\rightarrow L}^{-1}\right)}
\left(\frac{1}{\mu_R}-\frac{1}{\mu_L}\right),
\label{A1}
\\
% A_2&=\frac{ge^4v^3}{12\hbar c^2\mu^4\left(\tau_{L\rightarrow R}^{-1}+\tau_{R\rightarrow L}^{-1}\right)^2}.
A_2 &= \frac{ge^4v^3}{24\hbar c^2 \left(\tau_{L\rightarrow R}^{-1}+\tau_{R\rightarrow L}^{-1}\right)^2} \left(\frac{1}{\mu_L^4}+\frac{1}{\mu_R^4}\right).
\label{A2}
\end{align}
For weak electric and magnetic fields, the modifications
of the entropy given by Eqs.~(\ref{SdeltaN}) and (\ref{SEB}) are dominated by the terms of the lowest order in $\bE\cdot\bB$.
In a semimetal with asymmetric nodes ($\mu_L\neq\mu_R$ in equilibrium),
 the change in entropy is proportional to $\bE\cdot\bB$, and can be either positive or negative depending on the sign of $\bE\cdot\bB$ and on the relative chemical potentials of the two nodes.
In an (inversion-)symmetric Weyl semimetal, where $\mu_L = \mu_R$, the first-order terms $\propto\bE\cdot\bB$ in Eqs.~(\ref{SdeltaN})
and (\ref{SEB}) vanish, and the leading-order
field-induced correction to the entropy
is quadratic in $\bE\cdot\bB$ and always negative.

{\it Adiabatic dechiralisation.}
The decrease of entropy by external fields in a symmetric Weyl semimetal is clear intuitively:
the fields
transfer quasiparticles from one node to the other %(i.e. they induce a chirality imbalance)
and thus
make the system more ``ordered''. In an asymmetric semimetal, with two different chemical potentials at the nodes,
 the system becomes either more ``ordered''
or less ``ordered'' depending on the direction of transfer.

The effect of the external fields on the entropy of a Weyl semimetal is analogous to the change of entropy
of local magnetic moments in paramagnetic salts when they are placed in an external magnetic field~\cite{Klerk:demagnetization}.
In such systems, the magnetic moments become more ordered in an external field.
If the material is then removed from the field, the moments disorient and their entropy  increases.
If the system is adiabatically isolated,
this demagnetization process is accompanied by the absorption of heat from the other degrees of freedom, thus cooling the system,
which constitutes the essence of the adiabatic demagnetization method of achieving low temperatures~\cite{Klerk:demagnetization}.
In cases where the system is maintained in contact with a thermostat at a constant temperature, the demagnetization is accompanied
by an isothermal absorption of heat.
Similarly, when a symmetric Weyl semimetal is removed from parallel electric and magnetic fields, the chirality balance is restored,
and the system also absorbs heat, which may be detected experimentally. In the case of an asymmetric Weyl semimetal,
the heat may be either released or absorbed, depending on the alignment of the fields and on the relative chemical potentials.

In principle, Weyl semimetals are not adiabatically isolated in realistic experiments;
applying an electric field to the system (or creating gradients
of the chemical potential or temperature) leads to the generation of Joule heat. The amount
  of Joule heat generated per time (per unit volume) is given by
\begin{align}
	q\equiv \bE\cdot\bj=\sum_{i,j=L,R}\sigma_{ij}\left[\bE^2-\bE\cdot\bnabla\left(\frac{\mu_i}{T}\right)\right]
	\nonumber\\
	+\frac{ge^2}{4\pi^2\hbar^2c}(\delta\mu_L-\delta\mu_R)\bE\cdot\bB,
	\label{q}
\end{align}
where $\bj$ is the current density and $\sigma_{ij}$ is the response of
the current of quasiparticles at node $i$ to the electric field
acting on the electrons at node $j$ at zero magnetic field, with $\sigma=\sum_{i,j=L,R}\sigma_{ij}$ being
the conductivity of the system in zero magnetic field. If the fields are changing sufficiently slowly,
the Joule heat generated (per unit volume)
in a homogeneous system is given by integrating the rate (\ref{q}) with the quasistationary chemical potentials
given by Eqs.~(\ref{MuNi}) and (\ref{deltaN}) over the time of the experiment:
\begin{align}
	Q_J=\int\left[\sigma E^2+
	\frac{ge^4 v^3}{48\pi^2c^2\hbar}\left(\frac{1}{\mu_L^2}+\frac{1}{\mu_R^2}\right)
	\frac{(\bE\cdot\bB)^2}{\tau_{L\rightarrow R}^{-1}+\tau_{R\rightarrow L}^{-1}}\right]dt.
	\label{QJ}
\end{align}
The second term in the integrand in Eq.~\eqref{QJ} is the manifestation of the negative longitudinal
magnetoresistance arising from the chiral anomaly.

We emphasise, however, that generating Joule heat is independent of the thermal effects of dechiralisation.
Joule heat is the energy that
mobile charge carriers
receive from an external electric field, which is
then passed to phonons or to the other electrons, while the distribution
of electrons may be considered stationary.
 The effect of cooling or heating due to adiabatic
dechiralisation comes from redistributing electrons between the nodes, and it requires that a certain amount of work is done
on the system during the process of removing it from parallel electric and magnetic fields.
The two contributions to the total heat may be separated experimentally,
for example, using different time dependencies of the two effects and/or different
dependencies on external electric and magnetic fields, which we discuss
in more detail below.
This independence of the two contributions to the total heat
allows us to consider the effect of dechiralisation as if the system were adiabatically isolated.

In what immediately follows, we compute the change of the system's temperature during a sufficiently slow adiabatic dechiralisation
process, during which the fields $\bE$ and $\bB$ are switched off over a time scale longer than $\tau_{L\rightarrow R}$
and $\tau_{R\rightarrow L}$.
%% REPETITION: (which lie within dozens of picoseconds in realistic experiments).
Under this condition, the process may be considered as quasistatic and Eq.~(\ref{SEB}) may be applied at all times.
Utilising the identities
$\left(\frac{\partial S}{\partial T}\right)_{\bE\cdot\bB}\left[\frac{\partial T}{\partial \left(\bE\cdot\bB\right)}\right]_S
\left[\frac{\partial\left(\bE\cdot\bB\right)}{\partial S}\right]_T=-1$ and
$\left(\frac{\partial S}{\partial T}\right)_{\bE\cdot\bB}=C_{\bE\cdot\bB}/T$, where $C_{\bE\cdot\bB}$ is the
heat capacity of the system at constant $\bE\cdot\bB$, gives
\begin{align}
	\frac{\partial T}{\partial\left(\bE\cdot\bB\right)}=
	-\frac{T^2}{C_{\bE\cdot\bB}}\left[A_1 + 2 A_2\,\bE\cdot\bB+\ldots\right].
	\label{TEB}
\end{align}
Equation (\ref{TEB}) describes the change of the temperature of a Weyl semimetal due to adiabatic dechiralisation
when the external fields are changed.

Equation (\ref{TEB}) may be understood qualitatively as follows. If dechiralisation were isothermal,
the system would receive the amount of heat $\delta Q=T\left[\frac{\partial S}{\partial\left(\bE\cdot\bB\right)}\right]_T d(\bE\cdot\bB)$
from the environment upon changing the fields $\bE$ and $\bB$ infinitesimally. When the system is adiabatically isolated,
the same heat is taken from the phonons and/or from the kinetic energies of the quasiparticles with the heat capacity
$C_{\bE\cdot\bB}$. Utilising Eq.~(\ref{SEB}), this leads immediately to Eq.~(\ref{TEB}).

{\it Heat capacity.} Due to the transfer of particles between the nodes, parallel electric
and magnetic fields modify the heat capacity of a Weyl semimetal, which, in the experimentally
relevant regime of low temperatures $T\ll\mu_L,\mu_R$ is given by
\begin{align}
	C_{\bE\cdot\bB}=C_0-A_1\bE\cdot\bB-A_2\left(\bE\cdot\bB\right)^2+\ldots,
	\label{CEB}
\end{align}
according to Eq.~(\ref{SEB}), where $C_0$ is the heat capacity (per unit volume) of a Weyl semimetal
in the absence of external fields and the coefficients $A_1$ and $A_2$ are given by
Eqs.~(\ref{A1}) and (\ref{A2}).
In general, $C_0$ includes contributions from electrons and phonons and,
depending on the way the system's temperature is measured,
may also be affected by the thermometer with which the system may be in contact.
In the case of sufficiently strong disorder or weak magnetic
field, which allows one to neglect the quantisation of the quasiparticle states,
and for sufficiently low temperatures under consideration, the heat capacity
is dominated by the electrons and is given by
$C_0=\frac{g\pi^2(\mu_L^2+\mu_R^2)}{18(v\hbar^3)}$ (the generic case of arbitrary temperatures
and chemical potentials is considered in Supplemental Material~\cite{Supplemental}).

The heat capacity (\ref{CEB}) affects the dependence of the temperature change during adiabatic dechiralisation
on $\bE\cdot\bB$ [cf. Eq.~(\ref{TEB})]. Moreover, the dependence of the heat capacity
on external electric and magnetic fields, routinely measured in experiment,
presents a direct way to observe the manifestations of the chiral anomaly in the thermodynamic properties
of Weyl semimetals.

{\it Estimates.} The effect of adiabatic dechiralisation is strongest in systems
with asymmetric Weyl nodes, which may be expected generically in Weyl semimetals
with broken inversion symmetry, such as in SrSi$_2$~\cite{ZHasan:SrSi2}. From Eqs.~(\ref{Ni}), (\ref{deltaN}), (\ref{TEB})
and (\ref{CEB})
 it follows that both
the relative change of the system's temperature as a result of dechiralisation
and the relative change of the heat capacity in the presence of electric and magnetic fields may be estimated as
\begin{align}
	\frac{\delta T_{\bE\cdot\bB}}{T}\sim\frac{\delta C_{\bE\cdot\bB}}{C_0}\sim\frac{\delta N_{\bE\cdot\bB}}{N}
	\sim \frac{\hbar v^3e^2\tau}{\mu^3 c} \bE\cdot\bB.
	\label{deltaTT}
\end{align}
Here we have assumed that the energy difference between the Weyl nodes
is of the same order of magnitude as the chemical potentials $\mu$ at both nodes (measured from the energies of the nodes)
and $\tau=\min(\tau_{L\rightarrow R},\tau_{R\rightarrow L})$.
For $v=10^8\frac{cm}{s}$, $\mu=10$\,meV, $\tau=10$\,ps, $E=0.1$\,V/mm and $B=1$\,T, Eq.~(\ref{deltaTT}) gives
$\delta T/T\sim 0.65$.
The temperature change $\delta T$ may have either sign, depending on which node has the larger chemical potential
and on the alignment of the field.

%Assuming the electric and magnetic fields are directed along the offset of node $R$ from node $L$
%in momentum space, the system will heat as a result of adiabatic dechiralisation if $\mu_L> \mu_R$ and will
%cool otherwise.

Equation (\ref{deltaTT}) for the change of temperature neglects the Joule heat generated in the system.
 The increase of temperature
due to Joule heating over time $t$ may be estimated, using Eqs.~(\ref{QJ}), (\ref{deltaN}) and (\ref{Ni}), as
\begin{align}
	\frac{\delta T_J}{T}\sim \frac{\sigma E^2}{TC_0}t
	+\left(\frac{\delta N_{\bE\cdot\bB}}{N}\right)^2\left(\frac{\mu}{T}\right)^2\frac{t}{\tau}.
	\label{QJestimate}
\end{align}
The effect of Joule heat is suppressed in the limits of small fields, short times of the experiment and large temperatures
($T\sim \mu$).
However, even in cases where this effect is not negligible, it may be separated from the effects of adiabatic dechiralisation
through its dependency on the fields and on the duration of the change of the fields. For instance, in the case of an
asymmetric Weyl semimetal under consideration, the contributions of dechiralisation and of Joule heating
to the temperature change are, respectively, linear and quadratic in the magnetic field.

The change of the heat capacity of the system when changing external electric and magnetic fields
is insensitive to Joule heating and, in principle, may be easier to observe in experiment. As our estimates above show,
the relative change of heat capacity [cf. Eq.~(\ref{deltaTT})] is of order unity for
asymmetric Weyl semimetals for experimentally accessible parameters.

In the case of a symmetric Weyl semimetal, with $\mu_L = \mu_R$,
the manifestations of the chiral anomaly are weaker than in the asymmetric case: 	
\begin{equation}
\frac{\delta T_{\bE\cdot\bB}}{T} \sim
\frac{\delta C_{\bE\cdot\bB}}{C_0}\sim
- \left( \frac{\delta N_{\bE\cdot\bB}}{N} \right)^2 \sim - \frac{\hbar^2 v^6 e^4 \tau^2}{\mu^6 c^2} (\bE\cdot\bB)^2.
\label{EstimatesSym}
\end{equation}
Using the same parameters as for an asymmetric semimetal gives $\delta T_{\bE\cdot\bB}/T \approx -0.027$.
In Eq.~(\ref{EstimatesSym}) we neglected the phononic contribution to the specific heat,
which is valid in the limit of low temperatures $T \ll \mu (s/v)^{3/2}$, where $s$ is the speed of sound.  This limit is realised in most experiments on Weyl semimetals~\cite{Note1}.
We emphasise that in symmetric Weyl semimetals the temperature increase due to Joule heating, described by
Eqs.~(\ref{QJ}) and (\ref{QJestimate}), always exceeds the temperature effect of dechiralisation, described by Eq.~(\ref{EstimatesSym}).
The two contributions, however, may be separated, e.g., by varying the time of the experiment.

We have focussed in this paper on the case
of Weyl semimetals with sufficiently large chemical potentials,
relevant for almost all experiments to date~\cite{Note1}.
Disordered Weyl semimetal with vanishing chemical potentials,
possibly realised in certain iridate compounds~\cite{SleightRamirez:T4},
may exhibit stronger thermodynamic signatures of the chiral anomaly,
with $\delta C_{\bE\cdot\bB}\propto T\left(\bE\cdot\bB\right)^\frac{2}{5}$
and $	\left[\frac{\partial T}{\partial\left(\bE\cdot\bB\right)}\right]_S
\propto T^2 (\bE\cdot\bB)^{-3/5}$ in the limit of low temperatures (see Supplemental
Material~\cite{Supplemental}).

{\it Conclusion and outlook.}
We have considered the thermodynamic manifestations of the chiral anomaly in Weyl semimetals.
We have studied, in particular, the dependence of heat capacity on external fields and the phenomenon of adiabatic dechiralisation,
 during which a change of
the imbalance between concentrations of quasiparticles with different chiralities results
in a change of the temperature of the system and/or heat absorption/release.
  These effects may be used to identify
systems with Weyl quasiparticle dispersion, even in cases where transport experiments are difficult. An interesting question, which in our view deserves further exploration,
is the interplay of adiabatic dechiralisation
with hydrodynamic transport, including such intriguing phenomena as generating turbulence and
magnetic instabilities~\cite{Galitski:dynamo} in the electron liquid.

In closing, we note that the basic phenomenology of adiabatic dechiralisation is not limited to Weyl semimetals,
but may manifest itself in any system where the chirality imbalance, i.e.
valley polarisation, is controlled by an external parameter.
When this parameter is switched on or off, the system will change its temperature and/or absorb/release heat.
In particular, if the valleys are symmetric in equilibrium, valley polarisation decreases the entropy of the system,
which leads to cooling or an absorption of heat during dechiralisation.
	
Here, we have focussed on the case of a Weyl semimetal, where the polarisation is induced by the parameter $\bE\cdot\bB$, which defines the chiral anomaly. However, in Weyl and other materials a chiral valley imbalance may be induced in other ways, such as by applying circularly polarised light (as in the metal dichalogenides~\cite{Mak:control, Zeng:valley}) or by using a magnetic field aligned with a particular crystallographic direction\cite{Zhu:field-induced, MacNeill:breaking, Arora:valley}. Analogous changes to the temperature and heat capacity can be expected in these cases.

{\it Acknowledgements.}
We are obliged to Art Ramirez for insightful discussions and for suggesting the term ``adiabatic dechiralisation''.
%We also appreciate useful discussions with L.S.~Levitov.
	BS gratefully appreciates the hospitality of the Physics Department at the University of California Santa Cruz, where a part
	of this work was completed.
	BS was also supported as part of the MIT Center for Excitonics, an Energy Frontier Research Center funded by the U.S.~Department of Energy, Office of Science, Basic Energy Sciences under Award no.\ DE-SC0001088.
The work of Y.R. was supported in part by the  Increase Competitiveness Program of NUST MISIS, Grant No. K2-2017-085,
and by the Russian Foundation for Basic Research (projects 17-02-00323 and 17-52-50023)

%%%%%%%%%%%%%%%%%%%%%%%%%%%%%%%%%%%%%BIBLIOGRAPHY%%%%%%%%%%%%%%%%%%%%%%%%%%%%%%%%%%%%%%%%%%%%%%%%%%%%%%%%%%%%%%%%%%%%%%%%%%%%%%
%%%%%%%%%%%%%%%%%%%%%%%%%%%%%%%%%%%%%%%%%%%%%%%%%%%%%%%%%%%%%%%%%%%%%%%%%%%%%%%%%%%%%%%%%%%%%%%%%%%%%%%%%%%%%%%%%%%%%%%%%%%%%%%

%%%%%%%%%%%%%%%%%%%%%%%%%%%%%%%%%%%%%%%%%%%%%%%%%%%%%%%%%%%%%%%%%%%%%%%%%%%%%%%%%%%%%%%%%%%%%%%%%%%%%%%%%%%%%%%%%%%%%%%%%%%%%%%
%%%%%%%%%%%%%%%%%%%%%%%%SUPPLEMENTAL%%%%%%%%%%%%%%%%%%%%%%%%%%%%%%%%%%%%%%%%%%%%%%%%%%%%%%%%%%%%%%%%%%%%%%%%%%%%%%%%%%%%%%%%%%%

\newpage

\renewcommand{\theequation}{S\arabic{equation}}
\renewcommand{\thefigure}{S\arabic{figure}}
\renewcommand{\thetable}{S\arabic{table}}
\renewcommand{\thetable}{S\arabic{table}}
\renewcommand{\bibnumfmt}[1]{[S#1]}

\setcounter{equation}{0}
\setcounter{figure}{0}
\setcounter{enumiv}{0}

\onecolumngrid
\newpage
\begin{center}
	\textbf{\large Supplemental Material for \\
		``Adaiabatic dechiralisation and thermodynamics of Weyl semimetals''}
\end{center}
\vspace{2ex}
%\twocolumngrid

%%%%%%%%%%%%%%%%%%%%%%%%%%%%%%%%%%%%%%%%%%%%%%%%%Entropy%%%%%%%%%%%%%%%%%%%%%%%%%%%%%%%%%%%%%%%%%%%%%%%%%%%%%%%%%%%%%%%%%%%%%%%%%%%%%%%%%%%%%%%%%%%
\subsection{Entropy of a Weyl semimetal}
In this section, we compute the entropy of quasiparticles in a Weyl semimetal.
Because the total entropy is a sum of the quasiparticle contributions at the two nodes, it is sufficient to compute
the contribution of one node only. Therefore, in the calculation below we consider only the vicinity of one node
with unbounded Weyl dispersion without the spin and/or valley degeneracies and omit the node index.

In order to regularise divergencies in thermodynamic functions it is convenient to compute them relative to those at zero temperature $T=0$.
For instance, the grandcanonical potential $\Omega(T)$ (per unit volume) may be measured from the value
\begin{align}
	\Omega(0)=-\lim_{T\rightarrow 0}
	T\int\limits_{-\infty}^\infty\nu(\ve)\,d\ve \ln(1+e^{(\mu-\varepsilon)/T})
	=\int\limits_{-\Lambda}^{0}\frac{4\pi(\ve+\mu)^2\ve\, d\ve}{(2\pi v\hbar)^3},
\end{align}
where $\Lambda$ is the effective ultraviolet energy cutoff.
The grandcanonical potential at an arbitrary temperature $T$ is given by
\begin{align}
\Omega(T) &
%\equiv -T\int\limits_{-\infty}^\infty\nu(\ve)\,d\ve \ln(1+e^{(\mu-\varepsilon)/T})
=-T^4\int_{-\infty}^\infty \frac{(s+\mu/T)^24\pi ds}{(2\pi v\hbar)^3}\ln(1+e^{-s})
\nonumber\\
& =-4\pi T^4\int_0^\infty \frac{(s+\mu/T)^2ds}{(2\pi v\hbar)^3}\ln(1+e^{-s})-4\pi T^4
\int_{-\infty}^0 \frac{(s+\mu/T)^2ds}{(2\pi v\hbar)^3}\left(s+\sum\limits_{n=1}^\infty(-1)^{k+1}\frac{e^{ns}}{n}\right)
\nonumber\\
&=\Omega(0)-4\pi T^4\int_0^\infty \frac{(s+\mu/T)^2ds}{(2\pi v\hbar)^3}\ln(1+e^{-s})
-4\pi T^4\int_{-\infty}^0\frac{(s+\mu/T)^2ds}{(2\pi v\hbar)^3}\sum\limits_{n=1}^\infty(-1)^{n+1}\frac{e^{ns}}{n}
\nonumber\\
&=\Omega(0)-8\pi T^4\int_{0}^\infty\frac{s^2+(\mu/T)^2}{(2\pi v\hbar)^3}\sum\limits_{n=1}^\infty(-1)^{n+1}\frac{e^{-ns}}{n}=
\Omega(0)-\frac{T^4}{(v\hbar)^3}\left(\frac{7\pi^2}{360}+\frac{\mu^2}{12 T^2}\right),
\end{align}
which gives immediately the contribution of one Weyl node to the entropy (per unit volume)
\begin{equation}
	S=
	-\left(\frac{\partial\Omega}{\partial T}\right)_\mu
	=\frac{7\pi^2 T^3}{90(v\hbar)^3}+\frac{\mu^2 T}{6(v\hbar)^3}.
	\label{entropyOneNode}
\end{equation}

%%%%%%%%%%%%%%%%%%%%%%%%%%%%%%%%%%%%%%%%%%%%%%%%%%%%%%%%%%%%%%%%%%%%%%%%%%%%%%%%%%%%%%%%%%%%%%%%%%%%%%%%%%%%%%%%%%%%%%%%%%%%%%%%%%%%%%%%%%%%%%%%%%%%%%%%%%%%%%%%%%%%%%%%%%%%%%%
\subsection{Heat capacity}

When changing the temperature of a Weyl semimetal with two different Weyl nodes,
the quasiparticles get transferred from the vicinity of one node to the vicinity of the other, while
the total concentration
\begin{align}
	N=\sum_{i=L,R}g\frac{\mu_i^3+\pi^2\mu_i T^2}{6\pi^2\hbar^3v^3}
	\label{Ntotal}
\end{align}
of electrons remains constant.
In equilibrium, the chemical potentials at the nodes, when measured from the same value of energy, coincide, which requires
\begin{align}
	\mu_R=\mu_L+\Delta,
	\label{MuDifference}
\end{align}
where $\Delta$ is the difference of energies of the nodes.

Utilising Eqs.~(\ref{Ntotal}) and (\ref{MuDifference}) and the conservation of the total electron concentration, we obtain the modification
of the chemical potentials $\mu_L$ and $\mu_R$ (measured from the nodes):
\begin{align}
	\frac{d\mu_L}{dT}=\frac{d\mu_R}{dT}
	=-\frac{2\pi^2(\mu_L+\mu_R)T}{3\mu_L^2+3\mu_R^2+2\pi^2T^2}.
	\label{MuDerivativeT}
\end{align}
Using Eq.~(\ref{entropyOneNode}) for the contribution of one Weyl node to the entropy of the system
 and Eq.~(\ref{MuDerivativeT}), we arrive at the heat capacity of a Weyl semimetal in the absence of
external electric and magnetic fields
\begin{align}
	C_0=T\left(\frac{\partial S}{\partial T}\right)_{N=const}
	=g\frac{7\pi^2 T^3}{15(v\hbar)^3}+g\frac{\mu_L^2+\mu_R^2}{6(v\hbar)^3}T-g\frac{2\pi^2T^3(\mu_L+\mu_R)^2}{3(v\hbar)^3[2\pi^2 T^2+3(\mu_L^2+\mu_R^2)]}.
\end{align}
In the limits of sufficiently low and high temperatures the heat capacity is given by
\begin{align}
	C_0=
	\begin{cases}
		\frac{7\pi^2 gT^3}{15(v\hbar)^3},\ \ T\gg\mu_{L,R}\\
		\frac{\pi^2}{3}\nu T,\ \ T\ll\max(\mu_L,\mu_R)
	\end{cases}
\end{align}
where $\nu = g (\mu_L^2+\mu_R^2)/(2 \pi^2 \hbar^3 v^3)$ is the density of states of Weyl quasiparticles at the Fermi energy.
We emphasise that in the limit of low temperatures and chemical potentials, the heat capacity
of a Weyl semimetal may be dominated by the contribution of phonons
$C_{ph}\sim\frac{4\pi^2}{5s^3\hbar^3}T^3$, where $s$ is the speed of sound in the system.

%%%%%%%%%%%%%%%%%%%%%%%%%%%%%%%%%%%%%%%%%%%%%%%%%%%%%%%%%%%%%%%%%%%%%%%%%%%%%%%%%%%%%%%%%%%%%%%%%%%%%%%%%%%%%%%%%%%%%%%%%%%%%%%%%%%%%%%%%%%%%%%%%%%%%%%%%%%%%%%%%%%%%%%%%%%%%%%
\subsection{Rate equations for quasiparticle concentrations at different nodes}

In this section we derive the equations for the dynamics of the electron concentrations near the two nodes of a Weyl
semimetal. The concentrations may change with time as a result of two independent effects: the chiral anomaly and the
internodal scattering of quasiparticles by impurities.
Since the effect of chiral anomaly in disordered systems has already been considered in a number of papers, including that
in disordered systems [see, e.g., Ref.~\onlinecite{Burkov:AnomalyDiffusive}], we focus here on the part of the dynamics which
comes from the impurity scattering and assume that there are no external electric or magnetic fields.

The kinetic equations for the distribution functions $f_i(s\bp)$ of quasiparticles near the nodes of a Weyl semimetal
are given by
\begin{equation}
\label{eq:kin}
\partial_t f_{i}(s\bp)=\frac{2\pi}{\hbar}\nim\sum_{s^\prime}\int \left|\langle i^\prime s^\prime\bpp|U|i s\bp\rangle\right|^2[f_{i^\prime}(s^\prime\bpp)-f_{i}(s\bp)]\delta(\ve_{i^\prime s^\prime\bpp}-\ve_{is\bp})\frac{d\bpp}{(2\pi\hbar)^3}
\end{equation}
where indices
$i$ and $i^\prime$ label the nodes ($i,i^\prime=L,R$);
indices $s,s^\prime=c,v$ label the conduction $(c)$ and the valence $(v)$ bands at each node;
$|is\bp\rangle$ is the state of a quasiparticle with momentum $\bp$ (measured from the node) at node $i$ and band $s$;
$\varepsilon_{is\bp}$ is the energy of the respective quasiparticle; $\nim$ is the impurity concentration
and $U$ is the potential of one impurity.

The matrix element $\left|\langle\bpp,i^\prime|U|\bp,i\rangle\right|$ has a non-universal
dependence on the directions of the quasiparticle momenta $\bp$ and $\bp^\prime$, which depends on the microscopic
details of the model of the Weyl semimetal. However, the exact angular dependence will affect only the numerical
coefficient in the rate equations. In what follows we assume, for the sake of concreteness, that
\begin{equation}
\int\frac{d\Omega^\prime}{4\pi}|\langle i^\prime s^\prime\bpp |is\bp \rangle|^2=\frac{1}{2},
\label{AngDependenceExample}
\end{equation}
provided the scattering between the states $|is\bp\rangle$ and $|i^\prime s^\prime\bp^\prime\rangle$
is allowed by energy conservation, where $\frac{d\Omega^\prime}{4\pi}$ is the integration over the direction
of the momentum $\bp^\prime$ (measured from the node $i^\prime$).
This choice of the angular dependence of the matrix element corresponds, for example, to a Weyl semimetal
with the (disorder-free) Hamiltonian
\begin{equation}
H(\bk) = \frac{1}{2m} \left({k}_{z}^{2} - k_{0}^{2}\right)\hat\sigma_{z}
+ v \left({k}_{x}\hat\sigma_{x} + {k}_{y}\hat\sigma_{y} \right)+\frac{\Delta}{2k_0} k_z,
\label{eq:weyl_pair}
\end{equation}
where $\Delta\ll vk_0$ is the energy difference between the two nodes. In this model, the two nodes are located near the
momenta $\bk_\pm=(0,0,\pm k_0)$, taking into account the smallness of the energy difference $\Delta$.
Equation~(\ref{AngDependenceExample}) is a result of explicit integration over the direction of the momentum
$\bp^\prime=\bk-\bk_-$ (or $\bp=\bk-\bk_+$) in the limit $|\bp|, |\bp^\prime|\ll k_0$.

Equations~(\ref{eq:kin}) and (\ref{AngDependenceExample}) give the kinetic equation for the distribution
functions $f_i(\varepsilon)$ of quasiparticle energies $\varepsilon$ in the form
\begin{equation}
\partial_t f_{i}(\ve)=-\frac{\pi\nim}{\hbar}|U(2k_0)|^2\nu_{i^\prime}(\ve)[f_{i}(\ve)-f_{i^\prime}(\ve)],
\label{eq:kinE}
\end{equation}
where $U(2k_0)$ is the matrix element of the impurity potential at the momentum $2k_0$ approximately equal to the
separation between the nodes in momentum space and we have taken into account the smallness of $\bp$ and $\bp^\prime$
in comparison with this separation.
Multiplying both parts of Eq.~(\ref{eq:kinE}) by the density of states $g\nu_i(\ve)$
at node $i$ and integrating with respect to $\varepsilon$,
we arrive at the rate of change of the concentration $N_i=g\int \nu_i(\varepsilon)[f_{i}(\ve)-f^0(\ve)]d\ve$
of electrons at node $i$:
\begin{equation}
\partial_t N_i=-\frac{\pi g \nim}{\hbar}|U(2k_0)|^2\int \nu_{i}(\ve)\nu_{i^\prime}(\ve)[f_{i}(\ve)-f_{i^\prime}(\ve)]d\ve,
\label{Ndynamics}
\end{equation}
where $f^0(\ve)$ is the equilibrium distribution function.

For an arbitrary deviation of the quasiparticle distribution from equilibrium,
the right-hand side of Eq.~(\ref{Ndynamics}) is a non-linear function of the concentrations $N_i(\varepsilon)$.
However, for small deviations, the rate equations may be linearised in
the deviations
$\delta N_i(\varepsilon)=N_i(\varepsilon)-N_i^0(\varepsilon)$ of the concentrations from their
equilibrium values
$N_i^0(\varepsilon)$:
\begin{align}
\frac{d N_R}{dt}=&
\frac{\delta N_L}{\tau_{L\rightarrow R}}-\frac{\delta N_R}{\tau_{R\rightarrow L}},
\label{Dynamics1}
\\
\frac{dN_L}{dt}=&
-\frac{\delta N_L}{\tau_{L\rightarrow R}}+\frac{\delta N_R}{\tau_{R\rightarrow L}},
\label{Dynamics2}
\end{align}
where $\tau_{L\rightarrow R}^{-1}$ and $\tau_{R\rightarrow L}^{-1}$
are the scattering rates, respectively, from node $L$ to node $R$ and from node $R$ to node $L$. In the limits of low
and high temperatures the scattering rates are given by
\begin{align}
	\frac{1}{\tau_{i\rightarrow j}}=
	\left\{
	 \begin{array}{cc}
	 	\frac{\pi g }{\hbar}\nim|U(2k_0)|^2\nu_j^0,    & T\ll \mu_L, \mu_R\\
	 	\frac{3\pi g T^2}{5\hbar^4 v^3}\nim|U(2k_0)|^2, & T\gg \mu_L, \mu_R,
	 \end{array}
	\right.
	\label{rates}
\end{align}
where $\nu_i^0$ is the density of states at the Fermi level at node $i$. In most Weyl semimetals the momentum separation $2k_0$
between the nodes is of order of inverse atomic distances. At the respective scales, the Fourier transform $U(2k_0)$ of the
impurity potential is unaffected by screening and is, therefore, temperature-independent.

%%%%%%%%%%%%%%%%%%%%%%%%%%%%%%%%%%%%%%%%%%%%%%%%%%%%%%%%%%%%%%%%%%%%%%%%%%%%%%%%%%%%%%%%%%%%%%%%%%%%%%%%%%%%%%

\subsection{Dechiralisation at zero chemical potential}

In this section, we consider adiabatic dechiralisation and the effect of external fields on the heat
capacity in a Weyl semimetal with zero equilibrium chemical potential. Applying electric and magnetic fields
to such a system leads to finite chemical potentials $\delta\mu_L=-\delta\mu_R$ at the nodes. If the temperature
is smaller than these chemical potentials, $T\ll|\mu_L|$, the rate equations for the electron concentrations
are nonlinear in the concentrations. Using Eq.~(\ref{Ndynamics}) with the density of states
$\nu_i(\varepsilon)=\frac{\varepsilon^2}{2\pi^2\hbar^3v^3}$ at one node, we arrive at the rate equations in the presence
of electron and magnetic fields in the form
\begin{align}
	\frac{dN_L}{dt}=-\frac{dN_R}{dt}=\frac{g e^2}{4\pi^2 \hbar^2 c}\bE\cdot\bB
	-\frac{g\nim|U(2k_0)|^2}{10\pi^3\hbar^7v^6}\delta\mu_L^5.
\end{align}
When the fields $\bE$ and $\bB$ are time-independent, they lead to establishing the
stationary values of the chemical potentials given by
\begin{align}
	\delta\mu_L=-\delta\mu_R=
	\left[
	\frac{5\pi\hbar^5 v^6e^2}{2\nim|U(2k_0)|^2c}\bE\cdot\bB
	\right]^\frac{1}{5}
	\label{MuNonEq}
\end{align}
and, as a result, changes the heat capacity:
\begin{align}
	\delta C_{\bE\cdot\bB}\equiv\frac{g}{6v^3\hbar^3}(\delta\mu_R^2+\delta\mu_L^2)T
	\propto \left(\bE\cdot\bB\right)^\frac{2}{5}T.
	\label{CmuZeroLowT}
\end{align}
The adiabatic dechiralisation effect, characterised by the derivative
\begin{align}
 \label{app:dtdeb}
	\left[\frac{\partial T}{\partial\left(\bE\cdot\bB\right)}\right]_S
	\equiv-\frac{T}{C_{\bE\cdot\bB}}\left[\frac{\partial S}{\partial\left(\bE\cdot\bB\right)}\right]_T=-\frac{T^2}{C_{\bE\cdot\bB}}\frac{1}{6v^3\hbar^3}
    \frac{\partial(\mu_L^2+\mu_R^2)}{\partial \bE\cdot\bB}\propto T^2 (\bE\cdot\bB)^{-3/5},
\end{align}
which is singular at $\bE\cdot\bB=0$, may be particularly strong in this regime.

Equations (\ref{CmuZeroLowT}) and (\ref{app:dtdeb}) are valid for sufficiently low temperatures $T$,
exceeded by the non-equilibrium chemical potentials given by Eq.~(\ref{MuNonEq}). In the opposite limit of high temperatures
$T\gg\mu_{L,R}$, the dynamics of the quasiparticle concentrations at the nodes are described by
Eqs.~(\ref{Dynamics1}) and (\ref{Dynamics2}) with the rates given by the high-temperature limit of Eq.~(\ref{rates}).
The stationary chemical potentials in this limit are given by
\begin{equation}
   \delta\mu_{L,R}=\frac{5e^2}{\hbar c}\frac{\mathbf{E}\cdot \mathbf{B}}{4\pi^3}\frac{(\hbar v)^6}{T^4|U(2k_0)^2|\nim}.
\end{equation}
and the temperature effect of adiabatic dechiralisation is described by
\begin{equation}
\frac{\partial T}{\partial(\mathbf{E}\cdot \mathbf{B})}=-\frac{T^2}{C_{\mathbf{E}\cdot\mathbf{B}}}\left(\frac{5e^2}{4\pi^3\hbar c}\right)^2
 \frac{2 \mathbf{E}\cdot\mathbf{B} (\hbar v)^{12}}{\pi^4|U(2p_0)|^4 \nim^2 T^8}.
 \label{dTdEB}
\end{equation}
At intermediate temperatures $T\sim \mu_{L,R}$, Eq.~(\ref{dTdEB}) matches Eq.~(\ref{app:dtdeb}).

\end{document}